\documentclass[]{raa}            
\usepackage{graphicx,times}
\usepackage{natbib}

\providecommand{\bm}[1]{\mbox{\boldmath$#1$\unboldmath}}

\def\kms  {km~s$^{-1}$}

\begin{document}

   \title{C$_{2}$H, HC$_{3}$N and HNC Observations in OMC-2/3
}

 \volnopage{ {\bf 20xx} Vol.\ {\bf 9} No. {\bf XX}, 000--000}
   \setcounter{page}{1}

   \author{Qiang Liu
      \inst{1, 2}
   \and Ji Yang
      \inst{1}
   \and Yan Sun
      \inst{1}
   \and Ye Xu
      \inst{1}
   }

   \institute{Purple Mountain Observatory, Chinese Academy of Science, 2 West Beijing Road, Nanjing, Jiangsu 210008, China; {\it qiangliu@pmo.ac.cn}\\
        \and
             Graduate School of the Chinese Academy of Science,
             Beijing 100080, China
\vs \no
   {\small Received [year] [month] [day]; accepted [year] [month] [day] }
}

\abstract{ For the first time, the OMC-2/3 region was mapped in
$\rm{C_{2}H}$ (1-0), $\rm{HC_{3}N}$ (10-9) and HNC (1-0) lines. In
general, the emissions from all the three molecular species reveal
an extended filamentary structure. The distribution of $\rm{C_{2}H}$
cores almost follows that of the 1300 $\mu$m condensations, which
might suggest that $\rm{C_{2}H}$ is a good tracer to study the core
structure of molecular clouds. The core masses traced by HNC are
rather flat, ranging from 18.8 to 49.5 $M_{\odot}$, while present a
large span for those from C$_2$H, ranging from 6.4 to 36.0
$M_{\odot}$. The line widths of both HNC and $\rm{C_{2}H}$ look very
similar, and both are wider than that of $\rm{HC_{3}N}$. The line
widths of the three lines are all wider than those from dark clouds,
implying that the former is more active than the latter, and has
larger turbulence caused by winds and UV radiation from the
surrounding massive stars. \keywords{ISM: abundances
--- ISM: individual (Orion Molecular Clouds) --- ISM: molecules
--- stars: formation } }

   \authorrunning{Qiang Liu, Ji Yang, Yan Sun \& Ye Xu}            
   \titlerunning{C$_{2}$H, HC$_{3}$N and HNC Observations in OMC-2/3}  
   \maketitle

\section{Introduction}           
\label{sect:intro}

The Orion A molecular cloud, at a heliocentric distance of 450~pc
(\citealt{ge89}, it is worth to note that \citet{me07} measure the
parallax distance recently and give a value of 420 pc.), is one of
the nearest active high mass star-forming region. To the northern
end of Orion A, the $\int $-shaped OMC-2/3 region is regarded as one
of the best sites to study both ``cluster'' and triggered star
formation due to its near distance. Therefore OMC-2/3 aroused great
interest since its discovery and was comprehensively studied
recently at a variety of wavelengths and variety of molecular
species.

\citet{chi97} identified at least 21 compact 1300 $\mu $m dust
continuum condensations in OMC-2/3, 16 of them embedded in OMC-2 and
the other in OMC-3. They suggested that the condensations in OMC-3
with $L_{bol}$/$L_{smm}$ $<$ 70, are Class 0 objects and thus
represent an earlier stage of evolution. The 3.6~cm free-free
emission revealed 14 sources, of which seven sources coincide well
with the 1300 $\mu $m condensations, yet no relation was found
between the 3.6~cm radio continuum and 1300 $\mu$m (\citealt{re99}).
\citet{wi03} observed CO(1-0) toward this region and identified 9
protostellar outflows.

Dense cores of this region were well studied in many molecular
species, e.g. CS and $\rm{C^{18}O}$ (\citealt{ca95}), $\rm{NH_{3}}$
(\citealt{ce94}), $\rm HCO^{+}$ and CO (\citealt{aso00}).
The chemical evolution of OMC-2/3 was also
widely studied in variety of molecular species, in particular the
 complex molecule species, such as CH$_3$OH, HC$_3$N, CCS.
\citet{jo03} studied the astrochemistry of OMC-2/3 in
$\rm{H_{2}CO}$, $\rm{CH_{3}OH}$ etc., and
found a trend that hotter cores are more likely to have higher CO,
$\rm{H_{2}CO}$, $\rm{CH_{3}OH}$, and CS abundance. \citet{ta08}
observed $\rm{N_{2}H^{+}}$ ,$\rm{HC_{3}N}$ and CCS toward Orion A
and found that the N-bearing molecules seem to be more intense in OMC-2.
\citet{ta10} investigated Orion A in CCS, $\rm{HC_{3}N}$, DNC, $\rm{HN^{13}C}$
and detected CCS emission in OMC-3 for the first time. They also
proposed that star formation activity seems to be responsible for
the enhancement of $\rm{HC_{3}N}$ intensity.

We mapped OMC-2/3 region in $\rm{C_{2}H}$, $\rm{HC_{3}N}$, and HNC.
These Molecular lines were widely studied in dark clouds and were
commonly accepted as good tracers of physics condition and chemical
evolution of dense cores. $\rm{C_{2}H}$ was first detected in
interstellar medium by \citet{tu74}. Then it has been detected in a
variety of interstellar environments (\citealt{wo80},
\citealt{hu84}). Recently, \citet{be08} further revealed that
$\rm{C_{2}H}$ can be found from the earliest infrared dark clouds
(IRDCs) to the later evolutionary stage of ultra-compact HII regions
and explained that it got replenished at the core edges by elemental
carbon from CO being dissociated by the interstellar UV photons. Due
to its small rotation constant, $\rm{HC_{3}N}$ creates a bunch of
transitions and are easily detectable in molecular clouds
(\citealt{vanden83}). Moreover, the transitions are likely to be
optical thin, so it was also an excellent dense gas indicator.
However, its mechanism of formation is still not very clearly.
\citet{wo97} proposed that it was originated from the reaction
between CN and $\rm{C_{2}H_{2}}$. \citet{sz05} gave a pathway of
formation $\rm{HC_{3}N}$ from a C$_{3}$ carbon cluster and ammonia.
Observations of these molecular lines can help us understanding the
chemistry of $\rm{HC_{3}N}$. HNC are considered to be relevant to
the formation mechanism of cyanopolyynes (HC$_{2n+1}$N) in the
future work.

\section{Observations}
\label{sect:Obs}

The observations were made during 2010 March and July with the PMO
13.7~m millimeter-wave telescope at Delingha, China. The observation
center ($\alpha=05^{\rm h}35^{\rm m}26.71^{\rm s}$, $\delta=-05\dg
10'04''$, equinox=2000.0) was adopted from \citealt{chi97}, the
location of FIR 4, which associates with the strongest 3.6~cm
emission (\citealt{re99}). The HNC(1-0), $\rm{HC_{3}N}$(10-9) and
$\rm{C_{2}H}$(1-0) were mapped over a 10\arcmin~ by 23\arcmin~
region with a grid spacing of 60\arcsec. The detailed observation
log is listed in Table~1.

  A SIS receiver with a noise temperature 75-145~K (DSB)
was used. The back end was an Fast Fourier Transform Spectrometer
(FFTS) of 16384 channels with a bandwidth of 1000~MHz and effective
spectral resolution of 61.0~kHz (0.20 \kms). With the 1000~MHz
bandwidth, HNC(1-0) and $\rm{HC_{3}N}$(10-9) were received
simultaneously. The position-switch mode was used. The system
temperatures were about 200-300~K during the observations. The
pointing accuracy was checked by regularly observations of point
sources and was estimated to be better than 9\arcsec. The main beam
size was about 60\arcsec at 115~GHz. The typical on source time for
each position is about 5 minutes. The main-beam efficiency
$\theta_{mb}$ was estimated by comparing the radiation temperatures
of calibration sources S140, NGC2264 and Orion A with the NRAO 11~m
results (e.g. \citealt{hu84}, \citealt{mo76}).

All the data were reduced by using the GILDAS software. We make
linear baseline subtractions to most spectra. At the velocity
resolution of 0.2 \kms\, the typical rms noise level is about 0.2~K
in T$_{A}^{*}$.

\begin{table}

\bc

\begin{minipage}[]{100mm}

\caption[]{ The observed transitions and rest
frequencies}\end{minipage}

\small
 \begin{tabular}{ccccccccccc}
  \hline\noalign{\smallskip}
Molecular & Transition &  $\nu $  & relative intensity&$B_{0}$&
$\mu $ \\
   &  &   (MHz) & &(MHz)&
   (D)\\
  \hline\noalign{\smallskip}
$\rm{C_{2}H}$  & J=3/2$\rightarrow $1/2 F=1$\rightarrow $1 &  87284.38 &4.25& 43474& 0.8  \\
& J=3/2$\rightarrow $1/2 F=2$\rightarrow $1 &  87317.05 & 41.67&   \\
& J=3/2$\rightarrow $1/2 F=1$\rightarrow $0 &  87328.70 & 20.75&   \\
& J=1/2$\rightarrow $1/2 F=1$\rightarrow $1 &  87402.10 & 20.75&   \\
& J=1/2$\rightarrow $1/2 F=0$\rightarrow $1 &  87407.23 & 8.33&   \\
& J=1/2$\rightarrow $1/2 F=1$\rightarrow $0 &  87446.42 & 4.25&   \\
$\rm{HC_{3}N}$ & J=10$\rightarrow $9 &  90978.99 & &4549& 3.72 \\
HNC & J=1$\rightarrow $0  &  90663.59 & &45332& 3.05 \\
  \noalign{\smallskip}\hline
\end{tabular}
\ec
\tablecomments{0.86\textwidth}{references:(\citealt{tu74})}
\end{table}


\section{Result and Discussion}
\label{sect:Res}

\subsection{Spectra and maps}

Figure 1 shows the spectra of observation center. The intensity
scale is given in T$_R^*$ in all figures of this study. Here all six
hyperfine components of $\rm{C_{2}H}$ N=1$\rightarrow $0 were
presented. The line profiles of both HNC and HC$_3$N show a very
symmetry Gaussian profile at this position. Figure 2 shows the
velocity integrated intensity contour and grey-scale maps for the
strongest components J=3/2$\rightarrow $1/2 F=2$\rightarrow $1 of
C$_2$H, HC$_3$N and HNC. The integrated velocity ranges from 8 to 14
\kms\ for all observed lines. The 1300 $\mu$m cores identified by
\citet{chi97} were marked by pluses in panel a. It is evident that
the distribution of 1300 $\mu$m cores shows the most resemblance to
that of our $\rm{C_{2}H}$ cores. In general, the emissions from all
the three molecular species revealed an extended filamentary
structure. And more than one condensation was detected in both OMC-2
and OMC-3. One exception is HC$_3$N emission in OMC-3, which is
almost negligible and shows very marginal detection
($\sim$3$\sigma$). The location of the each core was derived from
the channel maps by eye and by applying a threshold of 5$\sigma$ in
two adjacent channels. To improve the signal-to-noise ratio, the
channel width was resampled to be 0.5 \kms\. Finally, we identified
at least eight $\rm{C_{2}H}$ J=3/2$\rightarrow $1/2 F=2$\rightarrow
$1 cores indicated by A1-A8, two $\rm{HC_{3}N}$(10-9) cores
indicated by B1 and B2, and seven HNC(1-0) cores indicated by
(B1-B7) in OMC-2/3. The location of each cores was marked by plus in
Figure~2.
 The size of the core is characterized by the nominal core radius R
after beam deconvolution, which was calculated by,
\begin{equation}
  R=[\frac{A}{\pi}-(\frac{HPBW}{2})^{2}]^{1/2},
\label{eq:LebsequeI}
\end{equation}
where A is the measured area within the contour of half peak
intensity. The derived location and size of each core were
summarized in Table 2, which also lists the corresponding Gaussian
fitting results for the main components of C$_2$H, HC$_3$N and HNC,
including the line center velocity, the line widths and the bright
temperature. The shape of most cores was elongated due to the
compression from surrounding medium. The radius of $\rm{C_{2}H}$
cores range from 0.12 to 0.17 pc with a mean value of 0.14 pc, while
range from 0.15 to 0.21 pc with a mean value of 0.17 pc for HNC
cores. The size of the two $\rm{HC_{3}N}$(10-9) cores is much
smaller, which is only 0.08 and 0.09 pc about half value of radius
of both HNC and C$_2$H cores. In OMC-2, the radius of C$^{32}$S
(2-1) core was found to be about 0.1 pc with resolution of 53\arcsec
(\citealt{ca95}), and that of CS(1-0) core was found to be 0.18 pc
with resolution of 35\arcsec (\citealt{ta93}). The discrepancy in
radius traced by different molecular species might reflect that
$\rm{HC_{3}N}$(10-9) and C$^{32}$S (2-1), perhaps
$\rm{C_{2}H}$(1-0), trace much denser region than HNC(1-0) and
CS(1-0).

\begin{figure}

   \includegraphics[width=60mm, angle=-90]{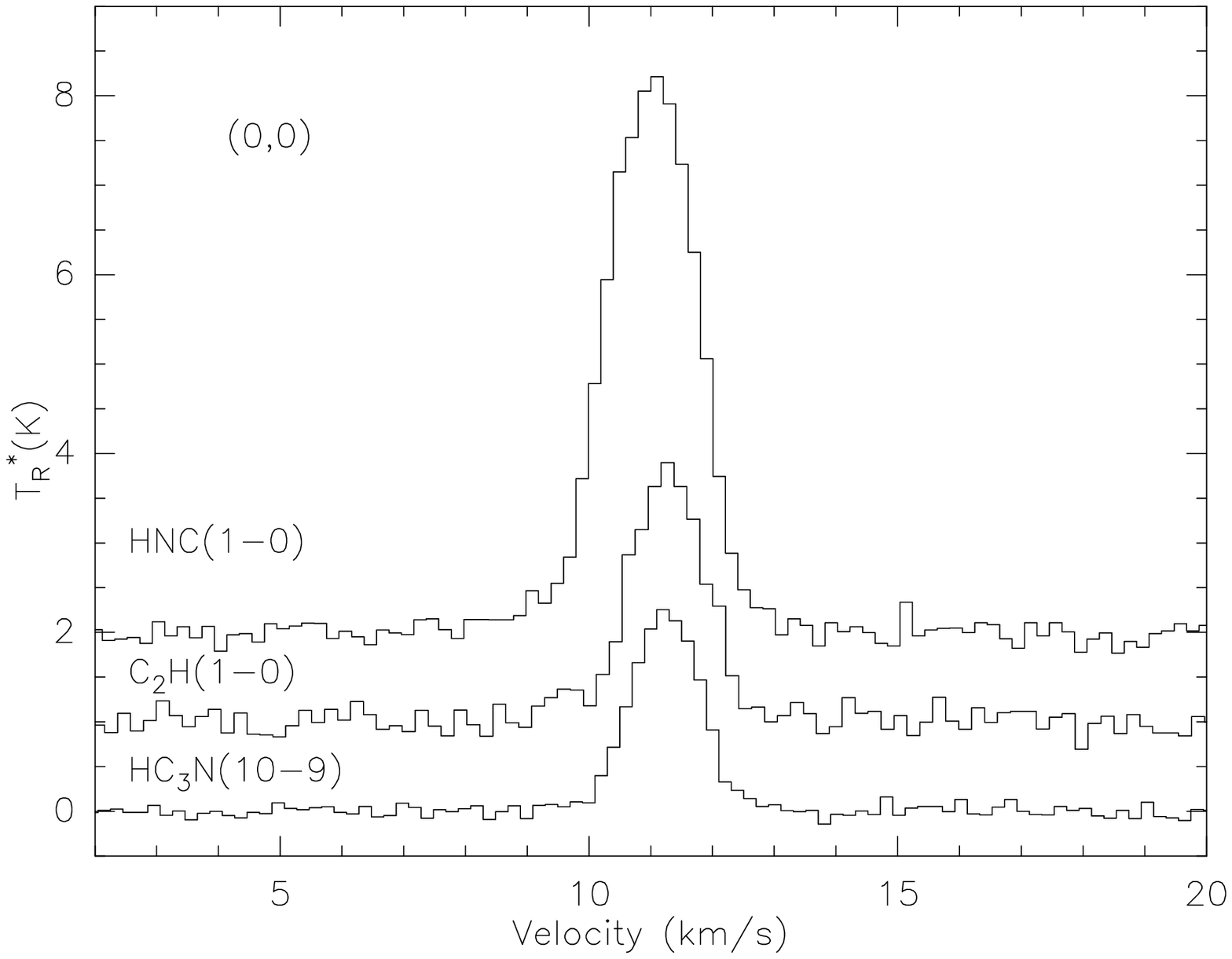}
   \includegraphics[width=60mm, angle=-90]{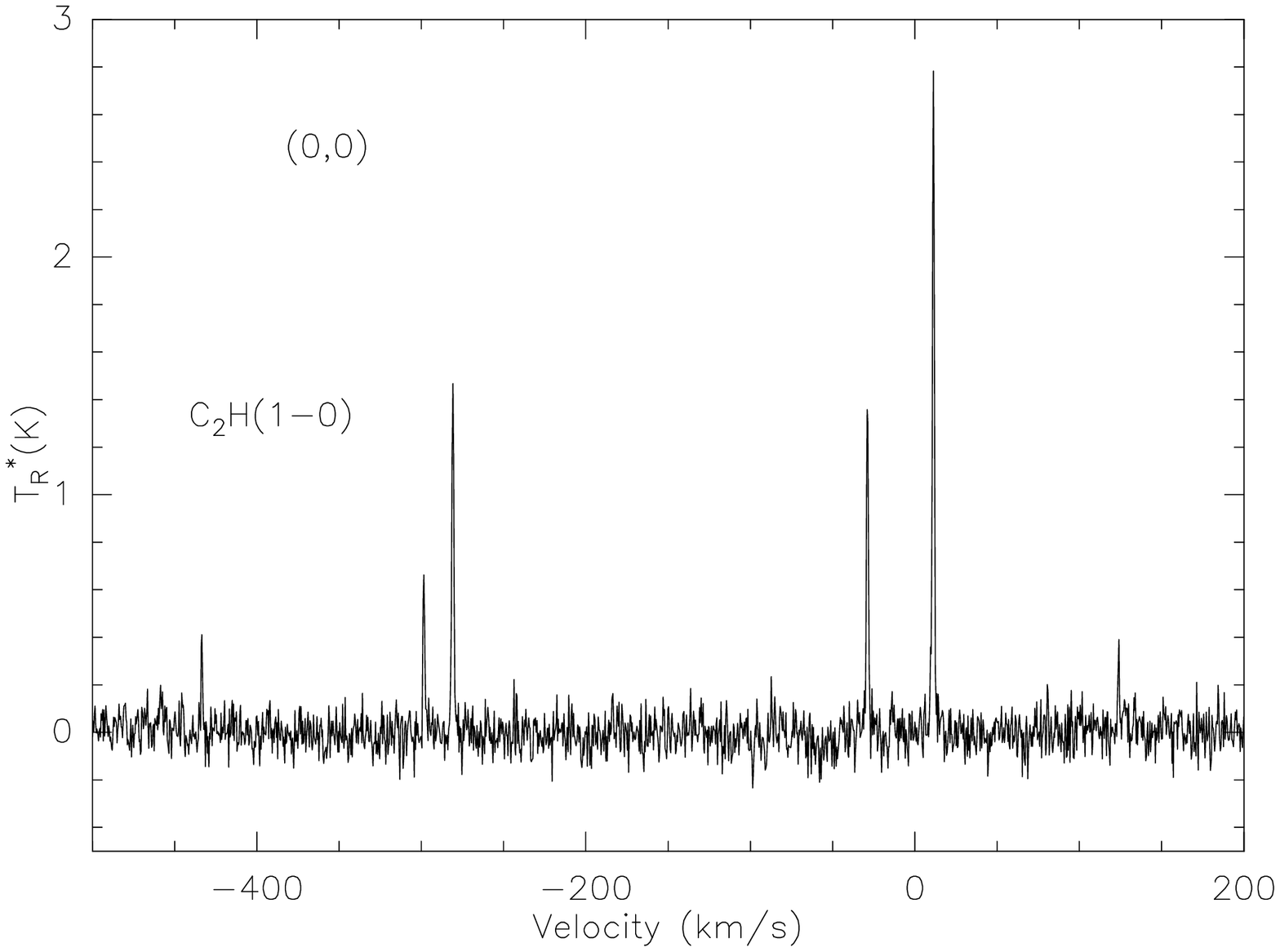}

   \caption{ Left: spetra towars (0$\arcsec$, 0$\arcsec$); Right: six hyperfine components of $\rm{C_{2}H}$ N=1$\rightarrow
 $0. }

  \label{Fig1}
   \end{figure}

\begin{figure}

\includegraphics[width=75mm, angle=-90]{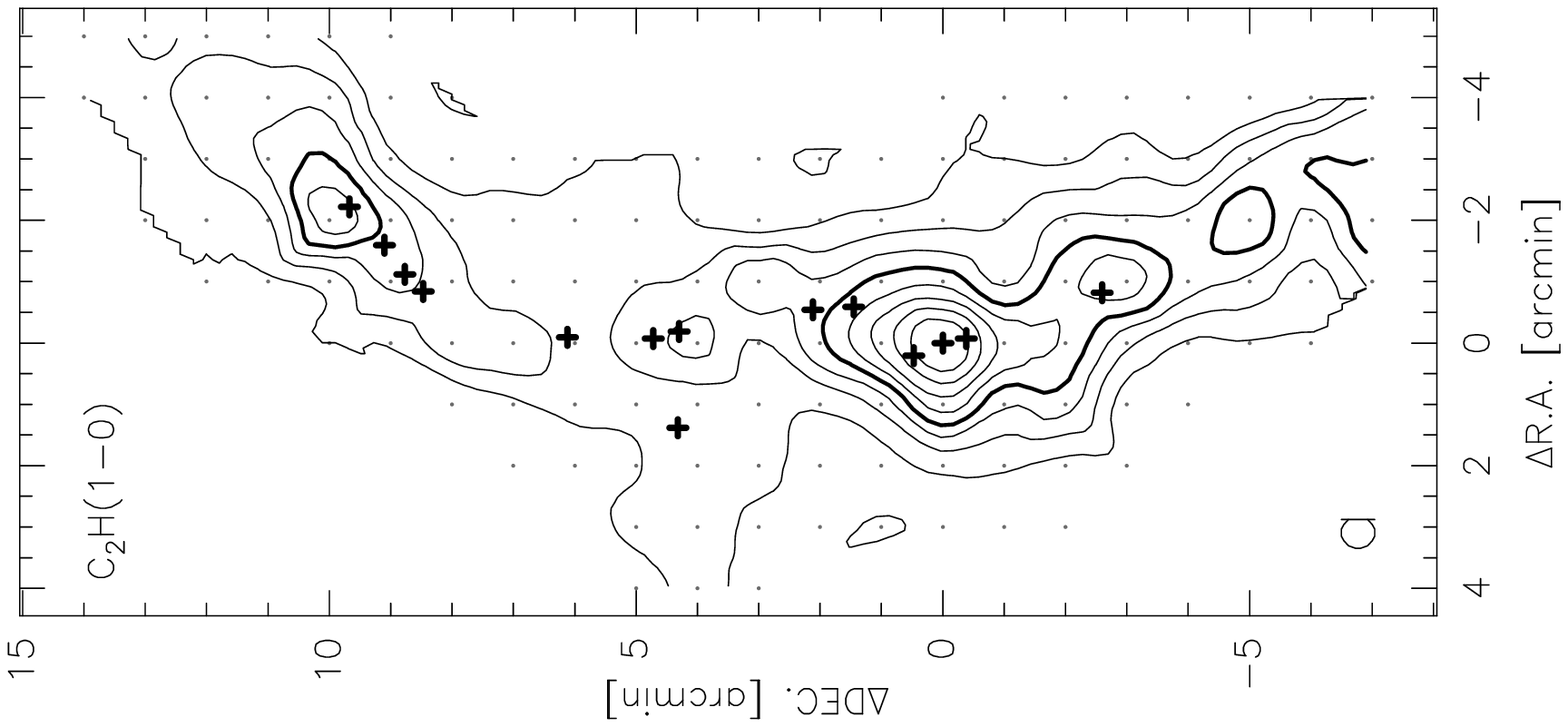}
\includegraphics[width=75mm, angle=-90]{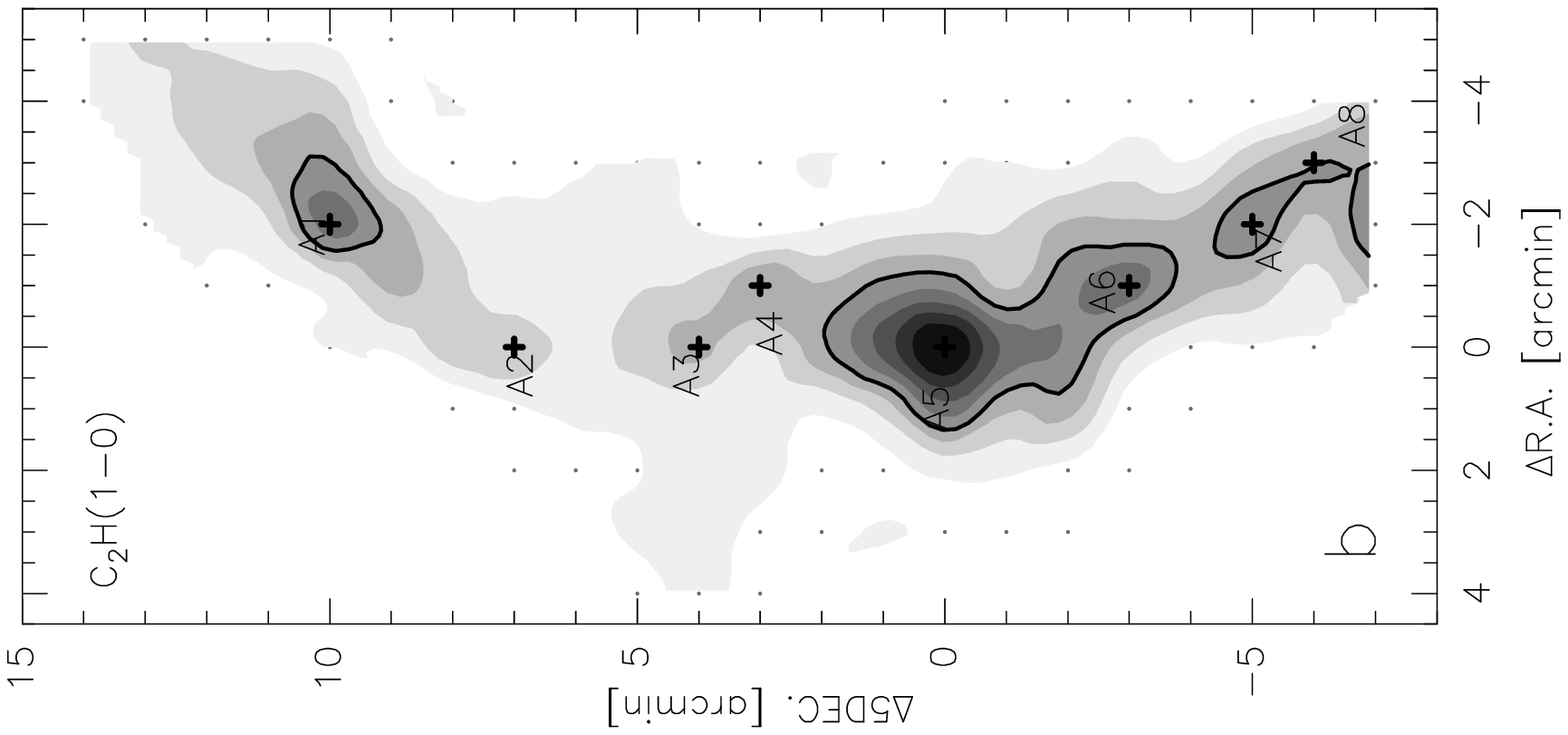}
\includegraphics[width=75mm, angle=-90]{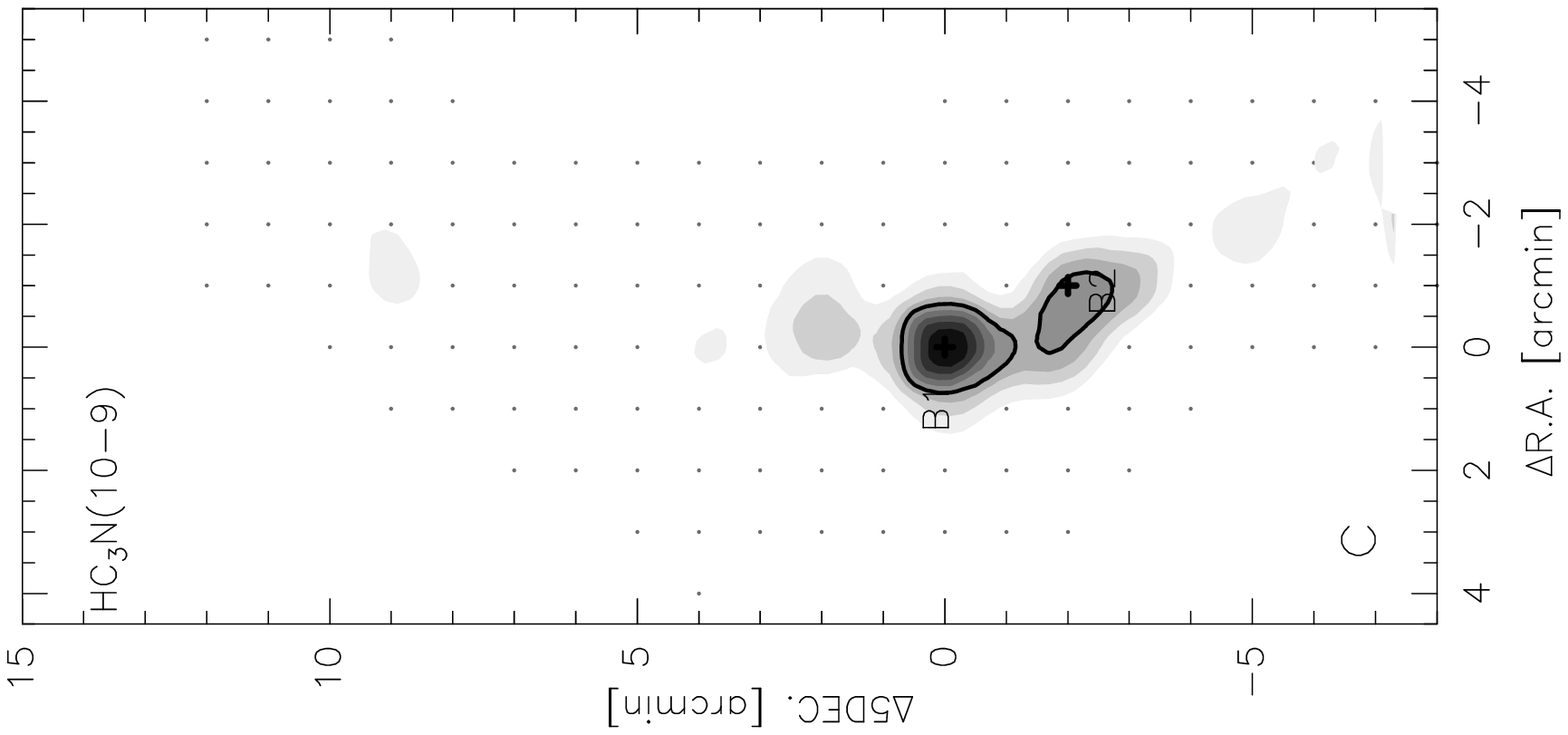}
\includegraphics[width=75mm, angle=-90]{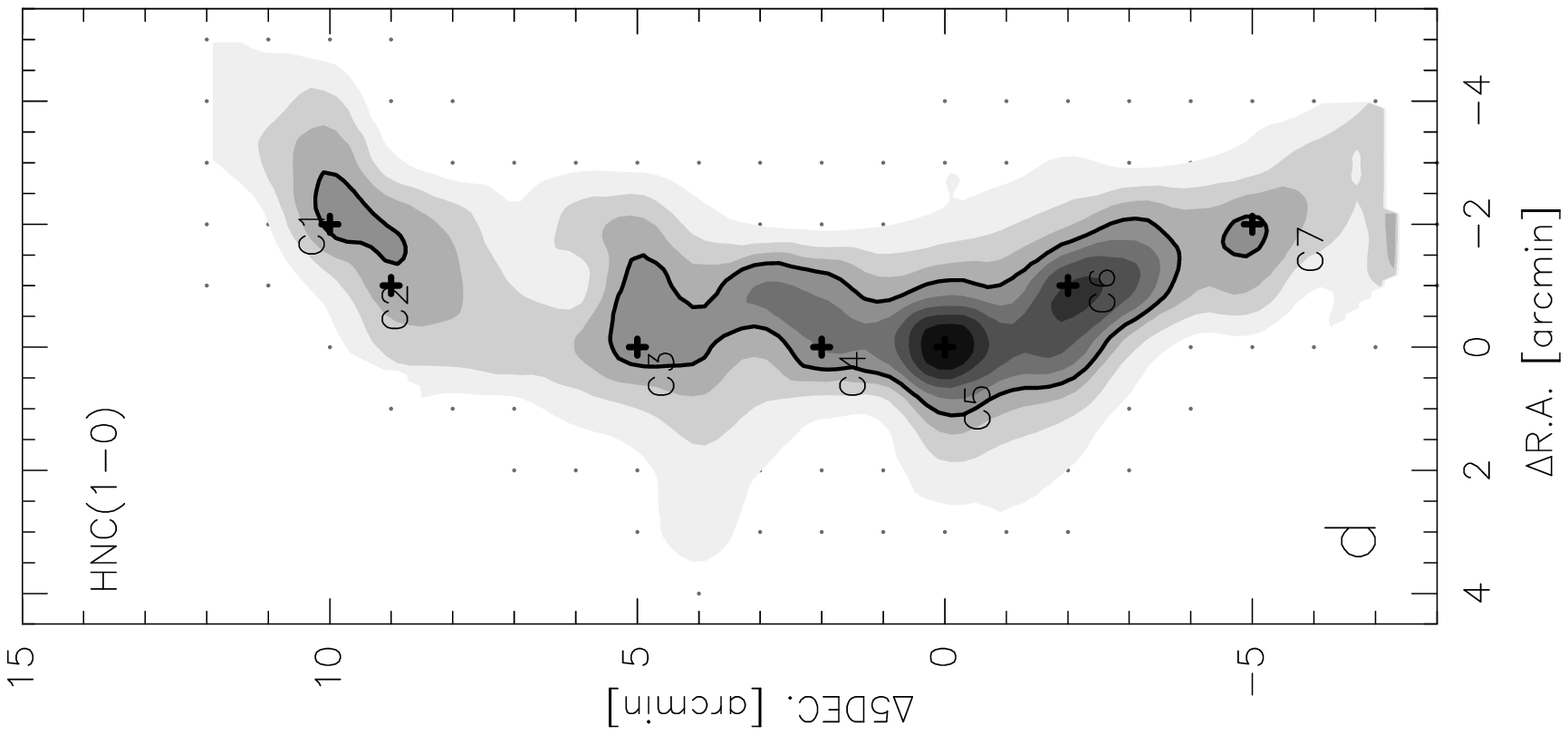}

\caption{a: velocity integrated intensity contour map of the
$\rm{C_{2}H}$ J=3/2$\rightarrow $1/2 F=2$\rightarrow $1, the contour
levels are 4.367 K \kms\ $\times $ (2,3,4,5,6,7,8,9). Pluses
represent the 1300 $\mu$m condensations identified by \citet{chi97}.
b-d: velocity integrated intensity contour and grey-scale maps of
the $\rm{C_{2}H}$ J=3/2$\rightarrow $1/2 F=2$\rightarrow $1,
$\rm{HC_{3}N}$(10-9), HNC(1-0). The contour levels are 4.367 K \kms\
$\times $ (2, 3, 4, 5, 6, 7, 8, 9), 1.785 K \kms\ $\times $ (3, 4,
5, 6, 7, 8, 9), 11.2 K \kms\ $\times $ (2, 3, 4, 5, 6, 7, 8, 9),
respectively. Pluses marked the positions of all cores identified in
this study.}
   \label{Fig2}
\end{figure}

\begin{table}

\bc

\begin{minipage}[]{100mm}

\caption[]{ Cores' parameters }\end{minipage}

\small
 \begin{tabular}{ccccccccccc}
  \hline\noalign{\smallskip}
No & offset &  offset  & R & R&
$V_{\rm LSR}$ & $\Delta $V & $T_{\rm mb}$ \\
   & (arcmin) & (arcmin) & (arcmin)&(pc)&  (\kms) & (\kms) & (K) \\
  \hline\noalign{\smallskip}
A1 & -2 &  10 & 1.34&0.17&10.9$\pm $0.04 &  1.8$\pm $0.08&1.4\\
A2 & 0 &  7 & 0.95&0.12&  10.9$\pm $0.04 &1.2$\pm $0.10&1.2\\
A3 & 0 &  4 &  1.18&0.15& 11.6$\pm $0.04 &1.4$\pm $0.12&1.1\\
A4 & -1 &  3 &0.92&0.12& 11.3$\pm $0.06  &1.7$\pm $0.15 &1.0\\
A5 & 0 &  0 & 1.32&0.17& 11.3$\pm $0.02  &1.4$\pm $0.04&2.8\\
A6 & -1 &  -3 & 0.94&0.12& 10.7$\pm $0.03  &1.5$\pm $0.08&1.6\\
A7 & -2 &  -5 & 1.00&0.13& 10.6$\pm $0.04  &1.5$\pm $0.11&1.5\\
A8 & -3 &  -6 & 1.01&0.13& 10.4$\pm $0.02  &0.85$\pm $0.07&1.8\\
\noalign{\smallskip}\hline \vspace{-2mm}\\
B1 & 0 &  0 &0.62&0.08&11.2$\pm $0.01&1.3$\pm $0.02&2.3\\
B2  & -1 & -2 &0.86&0.11&11.0$\pm $0.02&1.1$\pm $0.06&1.4\\
\noalign{\smallskip}\hline \vspace{-2mm}\\
C1 & -2 &  10 &1.17&0.15& 10.7$\pm $0.02 &  1.9$\pm $0.04&2.9\\
C2 & -1 &  9 & 1.35&0.18&10.9$\pm $0.01  & 1.5$\pm $0.03&3.3\\
C3 & 0 &  5 & 1.63&0.21&  11.2$\pm $0.01 &1.1$\pm $0.02&5.1\\
C4 & 0  &  2 & 1.27&0.17& 11.2$\pm $0.01 &1.5$\pm $0.02&4.2\\
C5 & 0 &  0 & 1.17&0.15& 11.0$\pm $0.01  &1.7$\pm $0.01&6.4\\
C6 & -1 &  -2 &1.35&0.18&  10.8$\pm $0.01  &1.3$\pm $0.02&6.3\\
C7 & -2 &  -5 &1.14&0.15&  10.4$\pm $0.01  &1.6$\pm $0.03&3.3\\
  \noalign{\smallskip}\hline
\end{tabular}
\ec

\end{table}

\subsection{Radial velocity and line width}
The velocities along the line of sight listed in Table 2 show a good
agreement among $\rm{C_{2}H}$, $\rm{HC_{3}N}$ and HNC cores. The
$V_{\rm LSR}$ ranges from 10.4 \kms\ to 11.6 \kms\ for $\rm{C_{2}H}$
(1-0) cores, and 10.4 \kms\ to 11.2 \kms\ for HNC(1-0) cores. The
$V_{\rm LSR}$ of the two $\rm{HC_{3}N}$(10-9) cores is all around
11.0 \kms.\ By comparing with the $V_{\rm LSR}$ derived from
C$^{18}$O (2-1) and C$^{32}$S (2-1) observations (Castetsa et al.
1995), we found that the $V_{\rm LSR}$ derived from different
molecular species are very coherent.

In the north-south direction, a velocity gradient across OMC-2 is
apparent in both HNC and C$_2$H, however, not the case for OMC-3
(see Table 2). Figure 3 shows the position-velocity diagram, for the
HNC (left) and $\rm{C_{2}H}$ (middle) along $\Delta$R.A. = -1, and
for $\rm{HC_{3}N}$ (right) along $\Delta$R.A. = 0 . The velocity
gradient in OMC-2 was obvious. Previously study also found that the
$V_{\rm LSR}$ increased gradually from south to north along the
Orion A filaments(e.g. \citealt{ba87}). One reasonable explanation
for its origin was the compression and acceleration from the
adjacent Orion OB I associations (\citealt{ba87}).

\begin{figure}

   \includegraphics[width=60mm, angle=-90]{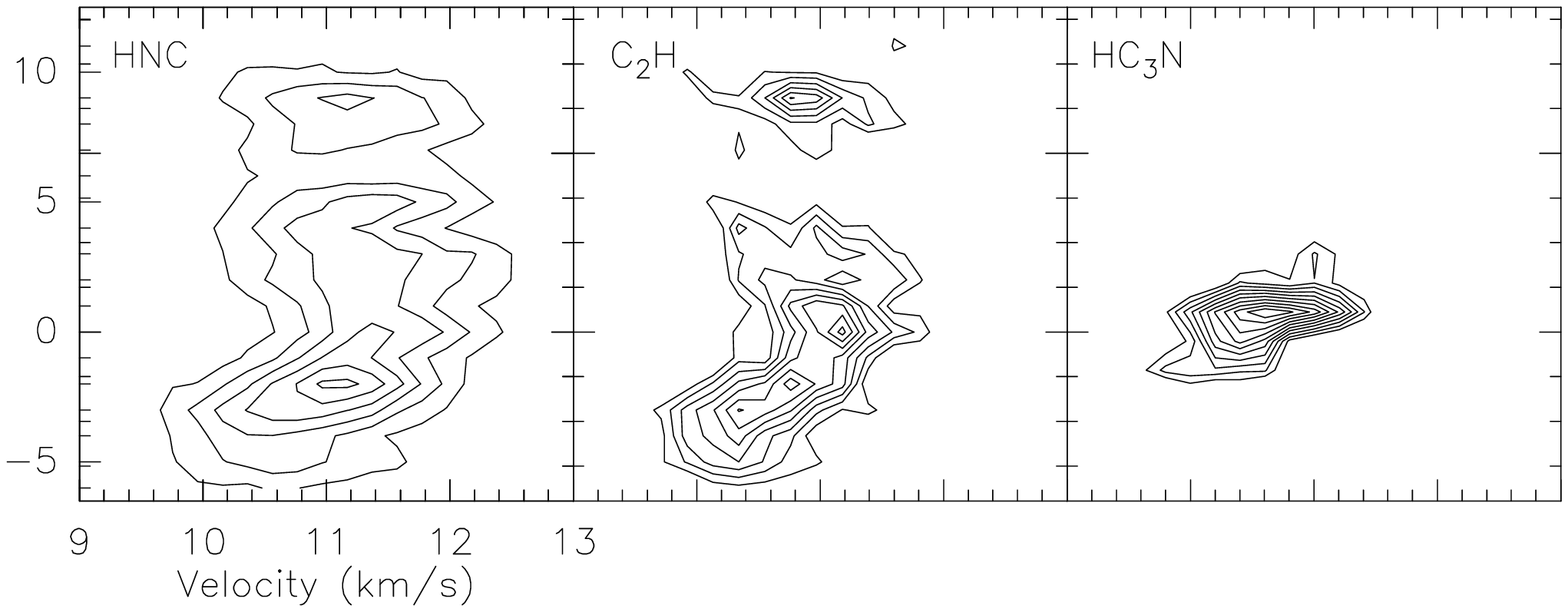}

\caption{The position-velocity diagram, for the HNC (left) and
$\rm{C_{2}H}$ (middle) along $\Delta$RA. = -1, and for
$\rm{HC_{3}N}$ (right) along $\Delta$RA. = 0 . }
 \label{Fig1}
   \end{figure}

The line widths of $\rm{C_{2}H}$(1-0), HNC(1-0), and
$\rm{HC_{3}N}$(10-9) range from 0.85 to 1.8 \kms,\ 1.1 to 1.9 \kms\
and 1.1 to 1.3 \kms,\ respectively. The line widths of both HNC and
$\rm{C_{2}H}$ look very similar, and both are wider than that of
$\rm{HC_{3}N}$. This indicates that $\rm{HC_{3}N}$ traces a cooler
region. As previously mentioned, $\rm{HC_{3}N}$ can trace much
denser and more center region of a cloud than HNC and $\rm{C_{2}H}$,
which might be the reason why HC$_3$N shows the narrowest line
width.

Many factors contribute to the line width of a specific molecular
specie. Among them, the contribution from thermal broadening is
about 0.2 \kms,\ so in our case non-thermal broadening is mainly
attributed to the line width. In dark clouds, the typical line width
is about 0.4-1.0~\kms\ for $\rm{C_{2}H}$(1-0) (\citealt{wo80}),
0.1-0.4 \kms\ for $\rm{HC_{3}N}$(10-9), and 0.5-0.7 \kms\ for
HNC(1-0) (e.g. in TMC-1; \citealt{ch84}), which are all much
narrower than those of OMC-2/3 region. This suggest that OMC-2/3 is
more active than dark clouds, and has larger turbulence caused by
winds and UV radiation from the surrounding massive stars.


\subsection{Column density and cores' masses}

We perform hyperfine structure (HFS) fitting to $\rm{C_{2}H}$ (1-0),
to estimate the excitation temperature T$_{\rm ex}$, which was given
as $T_{\rm ex}$=$T_{\rm ant}$ + $T_{\rm bg}$, and optical depth
$\tau_{TOT}$, which was given as $\tau _{\rm
 TOT}$=$\tau _{\rm main}$ / 0.4167 (\citealt{pa09}).

The total gas column density along the line of sight was calculated
under the assumption that each tracer is optically thin, in Local
Thermodynamic Equilibrium (LTE), and can be expressed as
(\citealt{sc86}),

\begin{equation}
  N=\frac{3k}{8\pi^3B\mu^2}\frac{e^{hBJ_{l}(J_{l}+1)/kT_{ex}}}{J_{l}+1}\frac{T_{ex}+hB/3k}{1-e^{-h\nu/kT_{ex}}}\int\tau_{\nu}d\nu,
\label{eq:LebsequeI}
\end{equation}
where B is the rotational constant, $\mu$ is permanent dipole
moment, and $J_{l}$ is the rotational quantum number of the lower
state in the observed transition (Table 1).
The excitation temperatures of $\rm{HC_{3}N}$ and HNC were assumed
to be 20 K in OMC-2 and 15 K in OMC-3, which were taken from $\rm
NH_{3}$ observations of \citet{wi99}.

The $\rm{H_{2}}$ column density N($\rm{H_{2}}$) was estimated by
assuming N($\rm{C_{2}H}$)/N($\rm{H_{2}}$)=$5.3\times10^{-9}$,
N($\rm{HC_{3}N}$)/N($\rm{H_{2}}$)=$1.3\times10^{-10}$, and
N(HNC)/N($\rm{H_{2}}$) =$5.3\times10^{-10}$ (\citealt{bl87}).
Core masses were calculated by equation, M= $\mu
m_{H}$N($\rm{H_{2}}$)$\times$($\pi R^{2}$), where $\mu$ the ratio of
total gas mass to hydrogen mass, is about 1.36.
 The viral mass $\rm{M_{vir}}$ was calculated by equation,
$\rm{M_{vir}}$ ($M_{\odot}$) = 210R(pc) $\delta $v(km / s). All
derived physical parameters were tabulated in Table 3.

An interesting result is that the core masses traced by HNC are
rather flat, ranging from 18.8 to 49.5 $M_{\odot}$. In contrast,
those traced by C$_2$H are steep, ranging from 6.4 to 36.0
$M_{\odot}$. On the whole, OMC-2 is more massive than OMC-3. The
average masses for $\rm{C_{2}H}$, $\rm{HC_{3}N}$ and HNC cores are
14.7$M_{\odot}$, 35.7$M_{\odot}$, 23.6$M_{\odot}$, respectively.
The core masses we derived here were strongly dependent on the
adopted abundance, therefore likely had a large uncertainty. The
average viral masses of $\rm{C_{2}H}$, $\rm{HC_{3}N}$ and HNC cores
are 41.5$M_{\odot}$, 23.6$M_{\odot}$, 52.7$M_{\odot}$, respectively.
$\rm{M_{vir}}$/M listed in Table 3 is inversely proportional to the
M, which is consistence with the conclusion of \citet{lo89} that low
mass clumps are more likely to deviate
 from viral equilibrium.

\begin{table}

\bc

\begin{minipage}[]{100mm}

\caption[]{ The cores' properties}\end{minipage}

\small
 \begin{tabular}{ccccccccccc}
  \hline\noalign{\smallskip}
No &   $T_{\rm ex}$ & $\tau _{\rm TOT}$&
N& N($\rm{H_{2}}$)&n($\rm{H_{2}}$) & M & $\rm{M_{vir}}$&$\rm{M_{vir}}$/M \\
   & (K) & &  $(cm^{-2})$ &$(cm^{-2})$  & $(cm^{-3})$& $M_{\odot}$ & $M_{\odot}$& \\
  \hline\noalign{\smallskip}
A1 &  5.1&1.2$\pm $0.33&$1.2\times10^{14}$ &$2.3\times10^{22}$&$2.2\times10^{4}$&31.6&64.0&2.0\\
A2 &  4.8&1.2$\pm $0.54&  $7.5\times10^{13}$ &$1.4\times10^{22}$&$1.9\times10^{4}$&9.6&30.1&2.7\\
A3 &  7.0&0.4$\pm $0.42&  $4.8\times10^{13}$&$9.1\times10^{21}$&$1.0\times10^{4}$&9.9&43.9&4.0\\
A4 & 8.9&0.2$\pm $1.62&  $4.2\times10^{13}$  &$7.9\times10^{21}$&$1.1\times10^{4}$&5.6&42.6&7.1\\
A5 & 13.9&0.4$\pm $0.15&   $1.4\times10^{14}$&$2.6\times10^{22}$&$2.5\times10^{4}$&36.0&49.7&1.2\\
A6 &  8.8&0.4$\pm $0.28&  $7.2\times10^{13}$  &$1.4\times10^{22}$&$1.9\times10^{4}$&9.6&37.6&3.7\\
A7 &  11.7&0.2$\pm $0.60& $5.7\times10^{13}$ &$1.1\times10^{22}$&$1.4\times10^{4}$&9.0&40.8&4.2\\
A8 &  13.5&0.2$\pm $0.46&  $4.1\times10^{13}$ &$7.7\times10^{21}$&$1.0\times10^{4}$&6.4&23.1&3.3\\
\noalign{\smallskip}\hline \vspace{-2mm}\\
B1 &   && $1.2\times10^{13}$ &$9.2\times10^{22}$&$18.6\times10^{4}$&27.9&21.8&0.8\\
B2 &  &&  $6.1\times10^{12}$ &$4.7\times10^{22}$&$6.9\times10^{4}$&26.9&25.4&0.9\\
\noalign{\smallskip}\hline \vspace{-2mm}\\
C1 &   && $1.3\times10^{13}$ &$2.5\times10^{22}$&$2.7\times10^{4}$&26.7&59.6&2.2\\
C2 &   && $1.2\times10^{13}$ &$2.3\times10^{22}$&$2.1\times10^{4}$&35.8&56.4&1.6\\
C3 &   && $1.1\times10^{13}$ &$2.1\times10^{22}$&$1.6\times10^{4}$&43.4&48.3&1.1\\
C4 &   && $1.2\times10^{13}$ &$2.3\times10^{22}$&$2.2\times10^{4}$&31.6&53.3&1.7\\
C5 &   && $2.2\times10^{13}$ &$4.2\times10^{22}$&$4.5\times10^{4}$&44.4&53.3&1.2\\
C6 &   && $1.7\times10^{13}$ &$3.2\times10^{22}$&$2.9\times10^{4}$&49.5&48.9&1.0\\
C7 &   && $9.8\times10^{12}$ &$1.8\times10^{22}$&$1.9\times10^{4}$&18.8&50.2&2.7\\
  \noalign{\smallskip}\hline
\end{tabular}
\ec

\end{table}

\section{summary}

We firstly mapped OMC-2/3 region in $\rm{C_{2}H}$(1-0),
$\rm{HC_{3}N}$ (10-9) and HNC (1-0) by using the PMO 13.7m telescope.
Our main results are summarized as following:

(1) The distribution of C$_2$H cores shows the most resemblance to that
of the 1300 $\mu$m condensations, which might suggest that $\rm{C_{2}H}$
is a good tracer to study the structure of molecular clouds.

(2) HC$_3$N shows the narrowest line width, meanwhile the widths of
both HNC and C$_2$H share a very similar distribution. In general,
the line width of the three observed line presented here is wider
than that of dark cloud, this might imply that OMC-2/3 is more
active than dark cloud, and has larger turbulence caused by winds
and UV radiation from the surrounding massive stars.

(3) The core masses traced by HNC are rather flat, ranging from 18.8
to 49.5 $M_{\odot}$, while, in contrast, those traced by C$_2$H are
steep, ranging from 6.4 to 36.0 $M_{\odot}$.

\normalem
\begin{acknowledgements}
We would like to thank the 13.7 m Observatory staff for their
support during the observation. This work was supported by the
Chinese NSF through grants NSF 11003046, NSF 11073054, NSF 10733030,
and NSF 10621303, and NBRPC (973 Program) under grant 2007CB815403.
\end{acknowledgements}

\label{lastpage}


\begin{thebibliography}{99}
\small \setlength{\itemindent}{-3mm} \setlength{\itemsep}{-0.5mm}
\setlength{\baselineskip}{4.5mm}
\bibitem[{Aso} {et~al.}(2000)] {aso00} Aso, Yoshiyuki, Tatematsu, Ken'ichi, Sekimoto, Yutaro, Nakano,
Takenori, Umemoto, Tomofumi, Koyama, Katsuji, Yamamoto, Satoshi
2000, APJS, 131, 465
\bibitem[{Bally} {et~al. }(1987)] {ba87} Bally, John; Lanber, William D., Stark, Antony A., Wilson, Robert
W. 1987, APJL, 312, L45
\bibitem[{Beuther} {et~al.}(2008)] {be08}
 Beuther, H., Semenov, D., Henning, Th., Linz, H. 2008, APJ, 675, 33
 \bibitem[{Blake} {et~al. }(1987)] {bl87} Blake, Geoffrey A., Sutton, E. C., Masson, C. R., Phillips, T.
G. 1987, APJ, 315, 621
\bibitem[{Castets\&  Langer
}(1995)] {ca95} Castets, A., Langer, W. D. 1995, A\&A, 294, 835
\bibitem[{Cesaroni\& Wilson }(1994)] {ce94} Cesaroni, R., Wilson, T. L. 1994, A\&A, 281, 209
\bibitem[{Chini} {et~al.}(1997)] {chi97} Chini, R., Reipurth, Bo., Ward-Thompson, D., Bally, J., Nyman, L.-A., Sievers, A., Billawala,
Y. 1997, APJ, 474, 135
\bibitem[{Churchwell} {et~al. }(1984)] {ch84} Churchwell, E., Nash, A. G., Walmsley, C.
M. 1984, APJ, 287, 681
\bibitem[{Genzel \& Stutzki}(1989)] {ge89} Genzel, Reinhard \& Stutzki,
Juergen 1989, ARA\&A, 27, 41
\bibitem[{Huggins} {et~al.}(1984)] {hu84} Huggins, P. J., Carlson, W. J., Kinney, A.
L. 1984, A\&A 133, 347
\bibitem[{Johnstone} {et~al.}(2003)] {jo03} Johnstone, D., Boonman, A. M. S., van Dishoeck, E.
F. 2003, A\&A, 412, 157
 \bibitem[{Loren}(1989)] {lo89} Loren, Robert B.
 1989,APJ, 338, 902
\bibitem[{Menten} {et~al.}(2007)] {me07} Menten, K. M., Reid, M. J., Forbrich, J., Brunthaler,
A. 2007 A\&A, 474, 515
\bibitem[{Morris} {et~al. }(1976)] {mo76}
 Morris, M., Turner, B. E. Palmer, P., Zuckerman, B. 1976, APJ, 205, 82
\bibitem[{Padovani} {et~al. }(2009)] {pa09} Padovani, M., Walmsley, C. M., Tafalla, M., Galli, D., M¨¹ller, H. S.
P. 2009, A\&A, 505 1199
 \bibitem[{Reipurth} {et~al.}(1999)] {re99} Reipurth, Bo, Rodr¨ªguez, Luis F., Chini,
Rolf 1999, AJ, 118, 983
 \bibitem[{Scoville } {et~al. }(1986)] {sc86}
 Scoville, N. Z., Sargent, A. I., Sanders, D. B., Claussen, M. J., Masson, C. R., Lo, K. Y., Phillips, T.
 G. 1986, APJ, 303, 416
  \bibitem[{ Szczepanski} {et~al. }(2005)] {sz05}
 Szczepanski, Jan, Wang, Haiyan, Doughty, Benjamin, Cole, Joseph, Vala,
 Martin 2005, APJ, 626, 69
\bibitem[{Tatematsu} {et~al.}(2010)] {ta10}
 Tatematsu, Ken'ichi, Hirota, Tomoya, Kandori, Ryo, Umemoto,
 Tomofumi 2010, arXiv1010.4939
\bibitem[{Tatematsu} {et~al.}(2008)] {ta08}
 Tatematsu, Ken'ichi, Kandori, Ryo, Umemoto, Tomofumi, Sekimoto,
 Yutaro 2008, PASJ, 60, 407
  \bibitem[{Tatematsu} {et~al.}(1993)] {ta93} Tatematsu, Ken'ichi et
 al. 1993, APJ, 404, 643
\bibitem[{Tucker} {et~al.}(1974)] {tu74}
 Tucker, K. D., Kutner, M. L., Thaddeus, P. 1974, APJ, 193, 115
\bibitem[{Vanden Bout} { et~al.}(1983)] {vanden83} Vanden Bout, P. A., Loren, R. B., Snell, R. L., Wootten, A. 1983, ApJ, 271, 161
\bibitem[{Williams} {et~al. }(2003)] {wi03} Williams, Jonathan P., Plambeck, R. L., Heyer, Mark
H. 2003, APJ, 591,1025
\bibitem[{Wilson} {et~al. }(1999)] {wi99}
 Wilson, T. L., Mauersberger, R., Gensheimer, P. D., Muders, D., Bieging, J.
 H. 1999, APJ, 525, 343
\bibitem[{Woon \& Herbst}(1997)] {wo97} Woon, David E.\& Herbst,
 Eric 1997, ApJ, 477, 204
\bibitem[{Wootten} {et~al.}(1980)] {wo80}Wootten, A., Bozyan, E. P., Garrett, D. B., Loren, R. B., Snell,
 R.L. 1980, APJ, 239, 844
\end{thebibliography}
\end{document}